
\input amstex
\documentstyle{amsppt}
\NoRunningHeads
\TagsAsMath
\TagsOnRight
\NoBlackBoxes
\font\special=cmr6
\topmatter
\title
Counting the local fields in SG theory.
\endtitle
\author
Feodor A. Smirnov\\
{\special Laboratoire de Physique Theorique
et Hautes Energies,}\\ {\special
4 place Jussieu,
Paris 75005, FRANCE }\\
{\special and}\\
{\special Steklov Mathematical Institute,}
\\{\special Fontanka 27, St. Petersburg 191011,  RUSSIA}
\endauthor
\abstract
In terms of the form factor bootstrap we describe
all the local fields in SG theory and check the agreement
with the free fermion case. We discuss the interesting structure
responsible for counting the local fields.
\endabstract
\endtopmatter
\def\a{\alpha}
\def\b{\beta}

\head
1. Introduction.
\endhead
In the present paper we solve the longstanding problem
of the description of all the local fields in SG theory in
terms of the form factor bootstrap approach. The results of
this paper add an important piece of information to the general
picture, so, this paper can be considered as an additional
chapter to the book [13]. It might be surprising that the
solution of the problem took such a long time, but it
required deeper understanding of the structure of SG form factors than
the author had when the book [13] was written. The most important
contributions to the field during the last several years
were added in the papers [5,14,6,15]
where the algebraic and the analitic properties of the form factors
were outlined.
It should be also said that
the necessity of a more general construction for the form factors
became clear to the author while thinking of the form factor
bootstrap for massless S-matrices [18] formulated in [4].

Let us formulate the problem. It is known that the form
factors of any local operator in the integrable field theory are
subject to the following axioms:

1. Riemann-Hilbert problem.
$$\align
& f(\b _1,\cdots,\b _i,\b _{i+1},\cdots ,\b _{n})S(\b _i-\b _{i+1})=
f(\b _1,\cdots,\b _{i+1},\b _i, \cdots ,\b _{n}),\\
& f(\b _1,\cdots ,\b _{n-1},\b _{n}+2\pi i)=
f(\b _n,\b _1,\cdots,\b _{n-1}), \tag {1} \endalign
$$

2. Residue condition
$$\align
&2\pi i \ \text{res}_{\b _n=\b _{n-1}+\pi i}
\ f(\b _1,\cdots,\b _{n-2} ,\b _{n-1},\b _{n}) = \tag {2}\\ &=
f(\b _1,\cdots,\b _{n-2})\otimes s_{n-1,n}\bigl(I-S(\b _{n-1}-\b_1)
\cdots S(\b _{n-1}-\b _{n-2})\bigr)
\endalign $$
where usual conventions are done [13], $S(\b)$ is a two particle S-matrix.
For the sake of simplicity we do not consider the problem
of bound states.

The classification of all the local operators in the theory is
equivalent to counting all the solutions to this infinite
system of equations. Let us ask ourselves what is the
best theory to attack the problem. The first possible
idea is that one has to consider the theory with the simplest
kind of S-matrix, for example, we can take
the scalar S-matrix (the theory in which particles
do not have internal degrees of freedom).
For such a model the equation (1) is satisfied trivially,
so the only problem left is to satisfy (2) which is
reduced to a certain recurrence relation for polynomials
in $e^{\pm\b _j}$. The problem was solved for several scalar S-matrices
in [3,8,9]. Notice that the S-matrices considered in [8,9] are
nontrivial ones.

In the present paper we take the opposite direction. We consider
the theory with a complicated S-matrix, which is
the soliton S-matrix for SG model:
$$ S =\int \bigl((\partial _{\mu}\phi)^2 +m^2\text{cos}(\beta \phi)
\bigr)d^2 x$$
In what follows we shall use only the renormalized
coupling constant:
$$\xi ={8\pi\over 8\pi -\b ^2}$$
In that case the equations (1) are very restrictive. Actually, they
can be viewed as a deformation of
level 0 Knizhnik-Zamolodchikov equations with trigonometric r-matrix.
The complete solution to this equation is available, moreover
the solutions are parametrized by a simple and understandable
structure (deformed cycles on hyperelliptic surface). So, to
solve the equation (2) one has to combine these solutions.
We shall see that this problem can be
solved completely, and thus all the local operators in SG theory
will be described. The space of operators obtained in this way
is complete because it coincides with the space of local operators
at the free fermion point. Moreover, we shall see that the
structure responsible for counting the fields does not
depend on the coupling constant. We shall comment on this
important point later.

It has to be said that the problem of description of all the local
operators in a situation similar to the SG one can be considered in
the framework of [6] (see also [11]).
In the terminology accepted in [6] the solution of the problem is
"a local operator is everything commuting with the
vertex operator of the second kind". We think that the
alternative answer given in the present paper is more
direct. It is important to understand the relation between these
two possibilities of answering the same question.

Let us describe the plan of the paper. The second and the third sections
are of introductory character: they describe the
necessary solutions to
KZ equations before and after the deformation.
The forth section is the central one where
the general construction
of the local fields is given. In the fifth section we explain the
agreement with the free fermion case and discuss the possibility of
taking into consideration the disorder type fields and the fields with
generalized statistics.
Finally, the last section contains the
discussion of further problems related to the subject of the paper.

\head
2. KZ equations for the trigonometric r-matrix on level 0.
\endhead

Consider the equations
$$b_i{d\over db_i}f(b_1,\cdots ,b_{2n})=
f(b_1,\cdots ,b_{2n})\bigl(\sum\limits _{j\ne i} r(b_i,b_j)\bigr)\tag {3}$$
where $ f(b_1,\cdots ,b_{2n})\in ((\Bbb{C}^2)^{\otimes 2n})^*$,
the trigonometric r-matrix $r(b_i,b_j)$ acts in the tensor product
of $i$-th and $j$-th spaces as follows:
$$r(b_i,b_j)={1\over 2}{b_i+b_j\over b_i-b_j}\sigma ^3_i\sigma ^3_j
+{\sqrt{b_ib_j}\over b_i-b_j}(\sigma ^1_i\sigma ^2_j +\sigma ^2_i\sigma ^1_j)$$
We
consider only the uncharged sector:
$$f(b_1,\cdots ,b_{2n})\Sigma ^3=0,\qquad
\Sigma ^3=\sum\limits _{i=1}^{2n} \sigma ^3_i$$
Moreover, we are interested in those solutions
annihilated by either of the two operators:
$$ \Sigma ^+_{\pm}=\sum\limits _{j=1}^{2n} (\sigma ^3_j)^j
\ b_j^{\pm{1\over 4}\sigma ^3_j}
\ \sigma ^+_j $$

Let us denote the set
of integers $\{1,\cdots ,2n\}$ by $U$. Different components
of the vectors $f$ are counted by the partitions $U=T\cup \bar{T}$,
where to be precise we require $1\in T$.
The solutions to these equations can be described as follows.

Consider
the function:
$$\align
&G_T(a_1,\cdots ,a_{n-1})=\\&=
{1\over\prod\limits _{1\ne i<j ,i,j\in T}(b_i-b_j)}
\text{det}\left|{1\over a_k-b_i}\prod\limits _{j\in T,j\ne i}
(a_k-b_j)\right|
_{k=1,\cdots ,n-1;\ i\in T,i\ne 1} \endalign $$
Then the solutions are given by
$$\align &
f_{\pm}^{\{\gamma _1,\cdots ,\gamma _{n-1}\}}(b_1,\cdots ,\b_{2n})_T
=\prod\limits_{i<j}(b_i-b_j)^{1\over 4}\tag {4}\\&\times
\prod\limits _{i\in T}b_i^{\pm{1\over 4}}
\prod\limits _{i\in \bar{T}}b_i^{\mp{1\over 4}}
\int\limits _{\gamma _1}{da_1\over\sqrt{P(a_1)}}\cdots
\int\limits _{\gamma _{n-1}}{da_1\over\sqrt{P(a_{n-1})}} G_T(a_1,\cdots
,a_{n-1})
\endalign $$
where $\{\gamma _1,\cdots ,\gamma _{n-1}\} $ are arbitrary cycles on the
hyperelliptic surface $f^2=\sqrt{P(a)}$ for $P(a)=\prod (a-b_j)$.

Let us count the number of different solutions. At the first glance
we have $2C_{2n-2}^{n-1}$ of them: $2$ is for $\pm$ and $C_{2n-2}^{n-1}$
is the number of choices of cycles. However this is greater than the
dimension of the space where $f_{\pm}$ live: $2(C_{2n}^{n}-C_{2n}^{n-1})$.
How to resolve this contradiction? There must be a linear dependence
between different solutions. Let us explain where it comes from.

The differentials which we have in (4) are obviously linear combinations
of those of the first and of the second kind, i.e. they do not
have residues on the surface. For the periods of such differentials
we have Riemann bilinear identity:
$$\sum\limits _{i=1}^{n-1} \left(
\int _{\a _i}\omega _1\int _{\b _i}\omega _2
 - \int _{\a _i}\omega _1\int _{\b _i}\omega _2\right)
=\sum\limits _{\text{poles}}\text{res}(\Omega _1 \omega _2)
\equiv\omega _1\circ\omega _2
$$
where $ \Omega _1 $ is a primitive function for  $\omega _1$;
$\a _i,\b _i$ is a canonic basis with the intersection numbers:
$$\a _i\circ \a _j=\b _i\circ \b _j=0,\ \a _i\circ\b _j=\delta _{ij}$$
It can be chosen as $\a _i=\delta _{2i}$, $\b _i=\sum _{k=1}^{i}\delta
_{2k-1}$ for $\delta _i$ drawn around the branching points $b_i$ and
$b_{i+1}$.

Consider the differentials from (4):
$$\zeta _{T,i}={1\over a-b_i}\prod\limits _{j\in T,j\ne i} (a-b_j)
{da\over\sqrt{P(a)}},\qquad i\in T\backslash 1 ,$$
for them we have
$$ \zeta _{T,i}\circ\zeta _{T,j}=0$$
because their poles and zeros cancel. For that reason one has
the following linear dependence between different solutions
$$\sum\limits _{i=1}^{n-1}
f_{\pm}^{\{\alpha _i,\beta _i
\gamma _3,\cdots ,\gamma _{n-1}\}}(b_1,\cdots ,\b_{2n})=0$$
for arbitrary $\{\gamma _3,\cdots ,\gamma _{n-1}\}$. So, the real
number of solutions is
$$2(C_{2n-2}^{n-1}-C_{2n-2}^{n-3})= 2(C_{2n}^{n}-C_{2n}^{n-1}) $$
which resolves the contradiction.

Dealing with the differentials of the first and of the second kind on
the hyperelliptic surface it is convenient to move the singularities
to the infinity. The basis of the differentials with singularities
at the infinity can be taken as
$$
\align
&\eta _p={a ^{p-1}\over \sqrt{P(a)}}da,\tag 5 \\
&\zeta _p= { 1 \over \sqrt{P(a)}}   \sum\limits _{k=p}^{2n-p}
(-1)^{k+p}(k-p)a ^{k-1}\sigma _{2n-p-k}(b_1,\cdots ,b_{2n})da
\endalign $$
where $p=1,\cdots ,n-1$ (recall that the genus of the surface is $n-1$),
$\sigma _{i}(b_1,\cdots ,b_{2n})$ are the elementary
symmetric polynomials.
This basis is canonic:
$$ \eta _i\circ\eta _j =  \zeta _i\circ\zeta _j =0,
\ \eta _i\circ\zeta _j=\delta _{ij} $$
The differentials from (4) can be rewritten in terms of (5).
Up to total derivatives one has
$$\align
\prod\limits _{i=1}^{n-1}{1\over\sqrt{P(a_i)}da_i}&\ G_T(a_1,\cdots ,a_{n-1})
\sim \\ &\sim
{1\over\prod\limits _{i\in T,j\in\bar{T}}(b_i-b_j)}
\text{det}\left|\zeta _{T,i}(a_k)\right|
_{k=1,\cdots ,n-1;\ i=1,\cdots ,n-1} \endalign $$
where
$$\zeta _{T,i}(a)={A_{T,i}(a)da\over \sqrt{P(a)}} $$
with the following polynomials:
$$A_{T,i}(a)=\prod\limits_{i\in T}(a-b_i)
\left[{\prod\limits_{i\in \bar{T}}(a-b_i)\over a^{n-i}}\right]_+   +
\prod\limits_{i\in \bar{T}}(a-b_i)
\left[{\prod\limits_{i\in T}(a-b_i)\over a^{n-i}}\right]_+ $$
where $[]_+$ means that only the polynomial part of the expression
in brackets is taken.
Considering $\zeta _{T,i}(a)$ as components of a vector one has
$$ \zeta _{T}(a)= \zeta (a) +C\eta (a) $$
where $C$ is certain $(n-1)\times (n-1)$ matrix. It is easy to
construct the differentials $\eta _{T}(a)$
dual to $\zeta _{T}(a)$ and thus
to obtain a symplectic matrix connecting  $\eta _{T}(a),\ \zeta _{T}(a)$
with $\eta (a),\ \zeta (a)$ from which $I$ and $C$ are two blocks.

Let us summarize the discussion of this section.
We see that the most important ingredient of the solutions to (3)
is the symplectic matrix composed of all the periods of
all the differentials of the first and of the second kind
$$\pmatrix P_1  \ & P_2 \\
           P_3 \ &  P_4
\endpmatrix $$
where $P_i$ are $(n-1)\times (n-1)$ matrices:
$$ \align
&(P_1)_{i,j}=\int\limits_{\a _i}\eta _j,\qquad
(P_2)_{i,j}=\int\limits_{\a _i}\zeta _j ,\\&
(P_3)_{i,j}=\int\limits_{\b _i}\eta _j,\qquad
(P_4)_{i,j}= \int\limits_{\b _i}\zeta _j
\endalign  $$

The solutions themselves can be considered as maps from
different polarizations of differentials ($\zeta _{T}(a)$ for any $T$
gives such a polarization) into different polarizations of
cycles (different half-bases of homologies).
Let us also mention that the above considerations are not very
different from those presented in [15] for the rational r-matrix,
for the reason that in our particular case (level 0
and special symmetry) the solution for trigonometric and rational
r-matrices are in one-to-one correspondence.

\head
3. Solutions to the deformed equations.
\endhead
Consider the equations
$$\align
& f(\b _1,\cdots,\b _i,\b _{i+1},\cdots ,\b _{2n})S(\b _i-\b _{i+1})=
f(\b _1,\cdots,\b _{i+1},\b _i, \cdots ,\b _{2n}),\\
& f(\b _1,\cdots ,\b _{n-1},\b _{2n}+2\pi i)=
f(\b _{2n},\b _1,\cdots,\b _{2n-1}), \tag {6} \endalign
$$
which can be regarded as a deformation of (3), the SG S-matrix is [17]
$$\align &S _{\xi}(\b)=
S_{\xi,0}(\b)
\widehat{S}_{\xi}(\b), \qquad
S_{\xi,0}(\b)=-\text{exp}\bigl(-i\int\limits _0^{\infty}
\frac {\text{sin}(k\b)\text{sh}({\pi-\xi \over 2}k) }
{k\text{sh}({\pi k\over 2})\text{ch}({\xi k\over 2}) }
\bigr),\\
&\widehat{S}_{\xi}(\b)  =
\frac { 1 } { \text{sh}{\pi \over \xi}(\b -\pi i)}
\pmatrix
 \text{sh}{\pi \over \xi}(\b -\pi i) &0 &0                              &0\\
 0    & \text{sh}{\pi \over \xi}\b    &-\text{sh}{\pi^2 i \over \xi}     &0\\
 0    & -\text{sh}{\pi^2 i\over \xi}    &\text{sh}{\pi \over \xi}\b    &0\\
 0 &0                             &0   &\text{sh}{\pi \over \xi}(\b -\pi i)
\endpmatrix ,\\
\endalign $$
Similarly to the classical case we consider only those
solutions of (6) which are invariant with respect to one of the two
quantum groups existing in the SG theory [12] which means
that they are annihilated by
$\Sigma ^3=\sum\limits _{i=1}^{2n} \sigma ^3_i$
and by either of the two operators:
$$ \Sigma ^+_{\pm}=\sum\limits _{j=1}^{2n} (\sigma ^3_j)^j
\ e^{\pm{\pi\over 2\xi}\b _j\sigma ^3_j}\ \sigma ^+_j $$
The variables exactly similar to $b_i$ from the previous
section are the following
$$b_i= e^{{2\pi\over \xi}\b _i}$$
The solutions to the equations (6) constitute a
linear space over the field of quasiconstants: symmetric $2\pi i$-periodic
functions, i.e. those depending on
$$B_i=e^{\b_i}$$ In what follows we shall use all this notations
($\b _j,\ b_j,\ B_j$).

Exactly as in the classical case, the most nontrivial part of
the solutions is described by the deformed matrix of the
periods of hyperelliptic differentials. The definition is as follows.

For two given polynomials $Q(a)$ and $L(A)$ (which can also
depend
on $b_j$
and on $B_j$ respectively as on parameters)
we consider the paring $\langle Q(a),L(A)\rangle$ defined by the integral:
$$
\langle Q(a),L(A)\rangle\equiv
\int\limits _{-\infty}^{\infty}\prod\limits _{j=1}^{2n}
\tilde{\varphi} (\a,\b _j)
\ Q(e^{{2\pi\over\xi}\a})L(e^{\a})
\ e^{{2\pi\over\xi}\a}
\ d\a \tag 7
$$
where
$$ \tilde{\varphi}(\a ,\b)=
\varphi(\a -\b)\ \text{exp}(-{1\over 2}({\pi\over\xi}+1) (\a+\b))$$
the function $\varphi$ is given by
$$
\varphi (\a)=C
\ \text{exp}\left(-2\int\limits _0^{\infty}
{{\text{sin}^2{k\a\over 2}\ \text{sh}{\pi+\xi\over 2}k}\over
{k\ \text{sh}{\xi k\over 2}\ \text{sh} \pi k}} dk\right) $$
where $C$ is certain constant [13].
The integral (7) is defined for $1\le\text{deg}(L(A))\le 2n-2 $,
and with a proper regularization [13] can be defined for an arbitrary
polynomial $Q(a)$.
We must understand the
polynomial $Q(a)$ as the one defining a deformed differential while
the polynomial $L(A)$ defines a deformed cycle. There is a canonic basis
of deformed differentials which is defined by the deformation of
the polynomials involved into (5):
$$\align
& R_p(a)=a^{p-1},\\ &
S_p (a)={1\over \tau-\tau ^{-1}}\sum\limits _{k=p}^{2n-p}
(-1)^{k+p}
(\tau ^{k-p} -\tau ^{p-k})
a ^{k-1}\sigma _{2n-p-k}(b_1,\cdots ,b_{2n})
\endalign $$
where $$\tau =e^{{2\pi^2 i\over\xi}}$$

The most important property of the deformed periods is that
they can be combined into a symplectic matrix. Namely,
consider the matrix
$$P=\pmatrix P_1\ &P_2\\P_3\ &P_4 \endpmatrix $$
where $P_i$ are the following $(n-1)\times (n-1)$ blocks:
$$ \align
&(P_1)_{i,j}=\langle R_j(a),A^{2i}\rangle,\qquad
 (P_2)_{i,j}=\langle S_j(a),A^{2i}\rangle,\\&
 (P_3)_{i,j}=\langle R_j(a),A^{2i-1}\rangle,\qquad
 (P_4)_{i,j}=\langle S_j(a),A^{2i-1}\rangle
\endalign  $$
It is shown in [16] that
$$ \pmatrix P_1\ &P_2\\P_3\ &P_4 \endpmatrix
   \pmatrix -P_4^t\ &P_2^t\\P_3^t\ &-P_1^t \endpmatrix=
\pmatrix X\ &0\\0\ &X\endpmatrix $$
where $X$ is the symmetric matrix composed of quasiconstants:
$$X_{i,j}=\int\limits _{-\infty}^{\infty}
{A^{2(i+j)} \over
\prod (A^2+B _j^2)}dA
$$
We could redefine the basis of the deformed cycles
combining them with quasiconstans in order to
eliminate $X$:
$$\a _i=A^{2i},\qquad \b _i =(X^{-1})_{ij}A^{2j-1}$$
We hope that $\a _i$ and $\b _i$ here (deformed cycles) will
not be confused with the rapidities and the integration variables.

The deformed period matrix allows a great simplification at the
free fermion point $\xi =\pi$. In fact, the calculation
of the period matrix at this point is not absolutely trivial,
one should be careful with the regularization of the integral (7)
explained in [13]. It can be shown that in the limit $\xi=\pi+\epsilon$,
$\epsilon\to 0$ the matrix $P$ has to be rescaled as follows:
$$
\tilde{P}=\pmatrix 1 &0\\0 &\epsilon\endpmatrix \ P$$
The matrix $\tilde{P}$ is finite in the limit:
$$\tilde{P}|_{\xi=\pi}=\pmatrix 0\ &I\\X\ &Y\endpmatrix \tag {8} $$
where
$$Y_{ij}=\sum\limits _{l=1}^{2n}
{1\over\prod^{\prime}_k (B _l^2-B _k^2)}
\sum\limits_{k=j}^{2n-j}
(k-j)B_l^{2i+2k-1}\sigma _{2n-j-k}(B_1^2,\cdots ,B_{2n}^2) $$
By a simple symplectic transformation of the deformed cycles this matrix can
be transformed into the unit matrix. This is an
important fact the meaning of which for
the SG model will be explained later.

It is possible to introduce the deformation of the integral
of a $k$-differential form over a $k$-chain:
$$\align &\langle Q(a_1,\cdots ,a_k),L(A_1,\cdots, A_k)\rangle=
\int\limits _{-\infty}^{\infty} d\a _1
\cdots
\int\limits _{-\infty}^{\infty} d\a _k
\prod\limits _{i=1}^{k}
\prod\limits _{j=1}^{2n}
\tilde{\varphi} (\a _i,\b _j)\\ &\times
Q(e^{{2\pi\over\xi}\a_1},\cdots ,e^{{2\pi\over\xi}\a_k})
L(e^{\a_1},\cdots ,e^{\a_k})
e^{{2\pi\over\xi}\Sigma\a_i}
\endalign $$
The polynomials $Q$ and $L$ are supposed to be antisymmetric; however,
later we shall take only $Q$ as antisymmetric then $L$ will
automatically antisymmetrize itself under the integral.

We would like to emphasize that contrary to the usual homologies which
constitute an exterior algebra over the ring of integers the
deformed homologies constitute an exterior algebra over the
ring of quasiconstants.

Let us return to the equations (6). Their solutions can be constructed
as follows:
$$\align
&f_{\pm}^{k_1,\cdots ,k_{n-1}}(\b_1,\cdots ,\b_{2n})_T=
\prod\limits_{i<j}\zeta(\b_i-\b_j)
\sum\limits_{T'\in U,\#T=n}{1\over\prod\limits_{i\in T,j\notin T}
\text{sh}{\pi\over\xi}(\b_i-\b_j)}\\&\times
e^{{n\over 2}\Sigma\b _j}
\ \tilde{f}_{\pm}^{k_1,\cdots ,k_{n-1}}(\b_1,\cdots ,\b_{2n})_{T'}
\ w(\b_1,\cdots ,\b_{2n}) _T^{T'}\endalign $$
where $\zeta(\b)$ is certain special function [13],
$w(\b_1,\cdots ,\b_{2n}) _T^{T'}$ is a special basis in $((\Bbb{C}^2)
^{\otimes 2n})^*$ [13]. The most interesting part of the
answer is the function
$\tilde{f}_{\pm}^{k_1,\cdots ,k_{n-1}}$
which is given by
$$
\tilde{f}_{\pm}^{k_1,\cdots ,k_{n-1}}(\b_1,\cdots ,\b_{2n})_{T}=
\langle F_T(a _1,\cdots a _{n-1}),
A_1^{k_1}\cdots A_{n-1}^{k_{n-1}})\rangle  \tag {9}
$$
where
$$F_T(a _1,\cdots a _{n-1})=\text{det}|A_{T,i}(a_j)|
_{(n-1)\times (n-1)}$$
the polynomials $A_{T,i} (a)$ are given by
$$ \align &
A_{T,i}(a)=
\prod\limits _{q\in T}(a-\tau b_q)
\sum\limits _{k=0}^{i-1}(1-\tau^{2i-2k})
a^{i-k-1}(-\tau)^{k}\sigma _k(b_{T'})
\\&\qquad +\tau ^{2i}
\prod\limits _{q\in T'}(a-\tau ^{-1}b_q)
\sum\limits _{k=0}^{i-1}(1-\tau^{2i-2k})
a^{i-k-1}(-\tau)^{k}\sigma _k(b_{T})
\endalign $$
where $b_T$ denotes the subset $\{b_j\}$ with $j\in T$.
There is another useful formula for
$\tilde{f}$ :
$$\tilde{f}_{\pm}^{k_1,\cdots ,k_{n-1}}(\b_1,\cdots ,\b_{2n})_{T}=
\text{det}\ \langle A_{T,i}(a), A^{k_j}\rangle _{i,j=1,\cdots , n-1}
$$

It is known that the functions $f$ satisfy (6) [13].
The arguments of the previous section concerning the
counting of solutions apply perfectly well to the deformed case.
Indeed, it can be shown that
$$A_T=S+CR$$
where $I$ and $C$ are blocks of a symplectic matrix. This fact
together with the deformed Riemann bilinear identity implies the
following linear dependence between the solutions:
$$\sum\limits _{i=1}^{n-1} (X^{-1})_{ij}
\tilde{f}_{\pm}^{2i,2j-1,
k_3,\cdots ,k_{n-1}}(\b_1,\cdots ,\b_{2n}) =0 \tag {10}$$
for arbitrary $ k_3,\cdots ,k_{n-1} $. Recall that the matrix $X$ is
composed of quasiconstants.

It can be shown that the solutions $f$ as functions of $\b _{2n}$
do not have other singularities for $0\le \text{Im}\b _{2n}\le 2\pi$
but the simple poles at $ \b _{2n}= \b _{j}+\pi i $. If we
impose this requirement generally and also require that the solutions
do not grow faster than $\text{exp}(x|\b _j|)$ for some $x$ as
$\b _j\to\pm\infty$
then the totality of solutions
to (6) is generated by
$$\text{exp}\bigl(\sum\limits_{l=-\infty}^{\infty}t_l(\sum e^{l\b_j})\bigr)
\ f_{\pm}^{k_1,\cdots ,k_{n-1}}(\b_1,\cdots ,\b_{2n}) \tag {11}$$
modulo the relation (10).
\head
4. Local operators in SG theory.
\endhead

In order to construct a local operator one has to find
a solution to the system composed of (6)
and
$$\align
&2\pi i \ \text{res}_{\b _{2n}=\b _{2n-1}+\pi i}
\ f(\b _1,\cdots,\b _{2n-2} ,\b _{2n-1},\b _{2n}) = \tag {12}\\ &=
f(\b _1,\cdots,\b _{2n-2})\otimes s_{2n-1,2n}\bigl(I-S(\b _{2n-1}-\b_1)
\cdots S(\b _{2n-1}-\b _{2n-2})\bigr)
\endalign $$
In the previous section we constructed all the solutions to (6),
so, our goal is to learn how to combine them in order to satisfy (12).
Let us explain the general idea about the organization of the
space of local fields in terms of form factor bootstrap.

We expect that there are infinitely many local operators which
are counted by the following data:
\item {1.} We fix the number of particles ($2m$) in the minimal form
factor. If we fixed $2m$ it means that for the given operator
the form factors with $2n$ particles vanish for $n<m$.

\item {2.} The minimal $2m$-particle form factor can be taken in infinitely
many possible ways due to a possible multiplication of the solution
to (6) by quasiconstant.

Keeping in mind the general formula (11) we write the following
generating function for the minimal $2m$-particle form factor:
$$\prod _{i<j}(e^{\b_i}+e^{\b_j})
\ \text{exp}\bigl(\sum\limits_{l=-\infty}^{\infty}t_l(\sum e^{l\b_j})\bigr)
\ f_{\pm}^{k_1,\cdots ,k_{n-1}}(\b_1,\cdots ,\b_{2n})_{T} \tag {13}
$$
modulo relation (10). This formula needs some comments. The
product $\prod _{i<j}(e^{\b_i}+e^{\b_j}) $ is put in order to
cancel the poles of  $f_{\pm}^{k_1,\cdots ,k_{n-1}}(\b_1,\cdots ,\b_{2n})$
which is needed for the form factor to be the minimal one:
the minimal form factor should not be
related to those with lower number of particles by
(12).
Different minimal form factors are obtained from (13) applying any
differential polynomial in $t_l$ and putting $t_l=0$ afterwards.
The times $t_l$
play a very different role for $l$ odd and even. The times $t_{2p+1}$ can be
identified with those related to the local integrals of motion, in particular
$t_1=z$, $t_{-1}=\bar{z}$
($z$ and $\bar{z}$ are usual euclidian coordinates) .
The multiplication of the form factors by
$\text{exp}(\sum\limits_{p=-\infty}^{\infty}t_{2p+1}(\sum e^{(2p+1)\b_j}))$
is always
possible because it does not spoil the residue condition.  The times $t_{2p}$
are far more nontrivial and, hence, more interesting: they do not
correspond to any local symmetry of the theory, but still they must be
considered. To summarize, in the formula (13) we have three structures counting
the minimal form factors: two bosonic (related to $t_{2p+1}$ and $t_{2p}$
respectively) and one fermionic (related to the counting by $k_1,\cdots
,k_{n-1}$). We shall show that to {\it every} minimal form factor (13)
a local operator can be related.

Let us consider the problem of calculation of the residue (12).
Generally the formulae for residue of
$f_{\pm}^{k_1,\cdots ,k_{n-1}}(\b_1,\cdots ,\b_{2n})_{T} $
can not be written in terms of the functions of the same kind.
There exists, however, one nice possibility which we are going to
describe.
It is useful to generalize the previous notations considering
the functions
$$ f_{\pm}^{L_n}(\b_1,\cdots ,\b_{2n})_{T} $$
where $L_n(A_1,\cdots ,A_{n-1})$ is an arbitrary polynomial whose
degree in any variable is between $1$ and $2n-2$. The generalization
corresponds to the
insertion of $L_n(A_1,\cdots ,A_{n-1})$ instead of
$A_1^{k_1}\cdots A_{n-1}^{k_{n-1}}$ in the formula (9).
The polynomial
$L_n(A_1,\cdots ,A_{n-1})$
can be taken as an antisymmetric one, but we prefer
to have it partly antisymmetric, namely, antisymmetric
in the variables $A_1,\cdots ,A_{k} $ for some given $k\le n-1$.
The polynomial
$L_n(A_1,\cdots ,A_{n-1})$ can also depend symmetrically on
$B_j$ as on parameters, we shall explicitly write this arguments
when needed.

The only situation where a good formula for the residue at
$\b _{2n}=\b _{2n-1}+\pi i$ exists is the following one
$$ \align
&L_n(A_1,\cdots ,A_{k},A_{k+1}\cdots ,A_{n-1} |
B_{1},\cdots ,B_{2n-2},B_{2n-1},B_{2n})|_
{B_{2n-1}=-B_{2n}\equiv B}=\\&=
\sum\limits_{i=1}^k(-1)^i\prod\limits_{l\ne i}(A_l^2+ B^2)
L^{\{i\}}_n(A_1,\cdots ,A_{k},A_{k+1}\cdots ,A_{n-1} |
B_{1},\cdots ,B_{2n-2}|B) \tag {14} \endalign $$
for some polynomials $L^{\{i\}}_n$. Provided the equation (14) holds we have
$$\align
&2\pi i \ \text{res}_{\b _{2n}=\b _{2n-1}+\pi i}
\ f_{\pm}^{L_n}(\b _1,\cdots,\b _{2n-2} ,\b _{2n-1},\b _{2n}) = \\ &=
f_{\pm}^{\tilde{L}_{n-1}^+}(\b _1,\cdots,\b _{2n-2})\otimes s_{2n-1,2n}+\\&+
f_{\pm}^{\tilde{L}_{n-1}^-}(\b _1,\cdots,\b _{2n-2})\otimes s_{2n-1,2n}
S(\b _{2n-1}-\b_1)
\cdots S(\b _{2n-1}-\b _{2n-2})
\endalign $$
where
$$\align
&\tilde{L}_{n-1}^{\pm}(A_1,\cdots ,A_{k-1},A_{k}\cdots ,A_{n-2}
|B_{1},\cdots ,B_{2n-2}|B)=\\&=
L^{\{k\}}_n(A_1,\cdots ,A_{k-1},\pm iB,A_{k}\cdots ,A_{n-2} |
B_{1},\cdots ,B_{2n-2}|B)  \tag {15}
\endalign $$
Obviously, the problem of finding a local operator with
the minimal form factor (13) will be solved if we find
the polynomials $L_m^m,L_{m+1}^m,L_{m+2}^m,\cdots $ such that:
\item {1.} For the given $n$ the integer
$k$ equals $n-m$, and the equation (14) holds.

\item {2.} The polynomials $\tilde{L}_{n-1}^{\pm}$ obtained from (15)
do not depend upon $B$. Moreover they are from the same sequence
of polynomials as $L_n^m$:
$$\align
&\tilde{L}_{n-1}^{m,\pm}(A_1,\cdots ,A_{k-1},A_{k}\cdots ,A_{n-2}
|B_{1},\cdots ,B_{2n-2}|B)=\\&=
\pm {B\over 2i}\ L_{n-1}^m(A_1,\cdots ,A_{k-1},A_{k}\cdots ,A_{n-2}
|B_{1},\cdots ,B_{2n-2}) \endalign $$

\item {3.}The initial condition is satisfied:
$$L_{m}^m(A_1,\cdots ,A_{m-1}
|B_{1},\cdots ,B_{2m}) =\prod\limits _{i<j}(B_i+B_j)
\prod\limits _{i=1}^{m-1}A_i^{k_i}$$

The polynomials which satisfy these conditions can be taken as
follows
$$\align
&L_n^m(A_1,\cdots ,A_{k},A_{k+1}\cdots ,A_{n-1} |
B_{1},\cdots ,B_{2n})=\\&=
\text{exp}(\sum\limits_{l=-\infty}^{\infty}t_l(\sum B_j^l))
{1\over  \prod\limits _{i<j}(B_i-B_j)}
\prod\limits _{i=1}^{n-m}A_i\ \prod\limits _{1\le i<j}^{n-m}
(A_i^2-A_j^2)
\ \prod\limits _{i=1}^{m-1}A_{n-m+i}^{k_i}\\&\times
\left|\matrix
  1,            &\cdots&\cdots,     &1                  \\
B_1^2           &\cdots&\cdots,   & B_{2n}^2             \\
  \vdots        &\vdots&\vdots       &\vdots             \\
B_1^{2(2n-m-1)} &\cdots&\cdots,    &B_{2n}^{2(2n-m-1)}   \\
H_1(B_1)        &\cdots&\cdots, &H_{1}(B_{2n})  \\
   \vdots       &\vdots&\vdots       &\vdots \\
H_{n-m}(B_1)    &\cdots&\cdots,    &H_{n-m}(B_{2n}) \endmatrix
\right| \tag {16} \endalign $$
where
$$H_i(B)=\text{exp}\bigl(-2\sum\limits_{p=-\infty}^{\infty} t_{2p}B^{2p}\bigr)
B^{2i-1}\prod\limits_{j=1}^{m-1}(A^2_{n-m +j}+B^2) $$
The formula (16) is the central formula of this work. It solves the
problem of constructing a local field from the minimal form factor.
Let us discuss this formula in some details. The times $t_{2p+1}$
enter it in rather trivial way. It is more interesting with
the times $t_{2p}$: they mix nontrivially with the deformed cycles.
As it has been said, we have added to SG two structures: one bosonic
($t_{2p}$) and one fermionic (deformed cycles). The formula (16)
defines a flow in this additional space. Notice also that
for $m=1$ and $t_{l}=0$ the formula simplifies a lot:
$$ L_n^1(A_1,\cdots ,A_{n-1}) =\prod\limits_{i=1}^{n-1}
A_i\prod\limits_{i<j}^{n-1}
(A_i^2-A_j^2)$$
This simple formula corresponds to the form factors
of the energy momentum tensor. It has been known for a long
time, but its simplicity misled the author preventing him
of finding the general formula (16).

There is also a question of the
uniqueness: whether the polynomials (16) are the only possible
for the given minimal form factor. It is not the case: we can
add to the operator with $2m$-particle minimal form factor
an arbitrary local operator whose minimal form factor has more
particles, the result being a local operator as well. So, the
problem is not that of the uniqueness, but that of the existence.
However, considering the problem of counting the local
operators we have to ignore the possibility of adding
a local operator with more particles in the minimal
form factor. Physically, the operator constructed via (16)
is characterized by the mildest possible ultraviolet behaviour
among those with the given minimal form factor.
\head
5. Agreement with the free fermion case and possible generalizations.
\endhead
We want to discuss the problem of completeness of our
list of local fields for SG model.
In modern language the completeness is related to the calculation of
certain characters. We shall take more simple route:
we shall consider the
agreement with what we have at the free fermion point.
Such an agreement would
imply the necessary character formulae.

Consider the case $\xi=\pi$ when SG model turns into the
free Dirac field:
$$S=\int(i\bar{\psi}\gamma _{\mu}\partial _{\mu}\psi+m\bar{\psi}\psi)d^2x $$
The solution to this model is given by the Fourier transform
$$\align &\psi (x_0,x_1)=\tag {17}
\\&=\int \left(e^{-ip_{\mu}(\b)x_{mu}}
\left(\matrix e^{{\b\over 2}}\\e^{-{\b\over 2}} \endmatrix\right)
a_-(\b)+e^{-ip_{\mu}(\b)x_{mu}}
\left(\matrix e^{{\b\over 2}}\\-e^{-{\b\over 2}} \endmatrix\right)
a_+^*(\b)\right)d\b \endalign $$
where $p_{\mu}(\b)=m(e^{\b}+(-1)^{\mu} e^{-\b})$.
The local field (note those of disorder type) for the
free fermion model have only one nontrivial form factor.
If we consider the neutral sector of the theory the form
factor  corresponds to the even number of particles
($2m$) the number of fermions being equal to the number of
antifermions, so the fields we are dealing with are of the
form
$$:\bar{D}_1(\bar{\psi})\cdots \bar{D}_m(\bar{\psi})
   D_1(\psi)\cdots D_m(\psi): \tag {18}$$
where $D_i,\bar{D}_i$ are arbitrary differential operators
with $\gamma$-matrix coefficients.
Not all of these operators are independent because we have to impose
the equations of motion, but there is now ambiguity
in their form factor description: the equations of
motion are taken into account by (17).

As usual we describe the spinor structure by subsets $T$
of those particles which are fermions (not antifermions).
The form factor corresponding to the operator of the
type (18) must be of the form
$$\align f(\b _1,\cdots & ,\b_{2m})_T=\tag {19}\\&
=\prod _{i<j,i,j\in T}\text{sh}{1\over 2} (\b _i-\b _j)
\prod _{i<j,i,j\in \bar{T}}\text{sh}{1\over 2} (\b _i-\b _j)
\ P_1(e^{\b _T})P_2(e^{\b _{\bar{T}}})\endalign $$
where $P_1$ and $P_2$ are two arbitrary symmetric Laurent polynomials
of $m$ variables. Not every operator which is local at the
free fermion point allows a local continuation for the general coupling.
The reason for that is in the fractional statistics of SG
solitons: only for $\xi=\pi$ they are real fermions. How to
describe those operators which do remain local for the nontrivial
coupling? The free fermion field is known to allow the infinite
dimensional symmetry algebra which is $U(sl(2))_{-1}$ [10], for
arbitrary coupling this algebra turns into the algebra $U(sl(2))_{q}$
($q=-\tau ^2$) of
nonlocal charges [2]. There are two finite-dimensional subalgebras
in this algebra, and we claim that only those local operators
invariant with respect to one of them can be locally continued
for the general coupling. In order to check the invariance one has
to make sure that the form factor (19) is annihilated by one of
two operators
$$ \Sigma ^+_{\pm}=\sum\limits _{j=1}^{2n} (\sigma ^3_j)^j
e^{\pm{1\over 2}\b _j\sigma ^3_j} \sigma ^+_j $$
which corresponds to the symmetry of the solutions of the form
factor equations which has been accepted in this paper.
It does not mean that other operators are really pathological,
they just have generalized statistics, we shall comment on them
later.

Consider the formula
$$P_1(B_1,\cdots ,B_m)\ P_2(B_{m+1},\cdots ,B_{2m}) \tag {20}$$
The space of polynomials (20) is
finite-dimensional over the ring of the Laurent polynomials,
symmetric in all the variables $B_1,\cdots ,B_{2m}$.
Taking this into account one realizes that the generating
function for the form factors
annihilated by  $ \Sigma ^+_{-}$
or $ \Sigma ^+_{+}$  can be written down as follows
$$
\text{exp}\bigl(\sum\limits_{l=-\infty}^{\infty}t_l(\sum e^{l\b_j})
\pm{1\over 2} (\sum\limits_{i\in T}\b_i -\sum\limits_{i\in \bar{T}}\b_i)\bigr)
\ P_{T}^{J}(e^{\b_1},\cdots ,e^{\b _{2m}}) \tag {21}$$
where $P_T^J$ are certain basic
polynomials, $J=1,\cdots ,C_{2n}^n -C_{2n}^{n-1}$ .
We do not write down an explicit formula for $P_J$ for the
economy of space.

The formula (21) looks very much similar to (13) the only
difference being in the way of counting from $1$ to
$C_{2n}^n -C_{2n}^{n-1}$: in (21) we count the polynomials
$P_T^J$ while in (13) we count $k_1,\cdots ,k_{m-1}$
(mod the relation (10)). However, the very simple form of the
period matrix at the free fermion point (8) allows to
establish the one-to-one correspondence between these two ways
of doing things. We do not present the explicit calculations
here because, to our mind, the situation is rather clear.

It is a proper place to discuss possible generalizations.
First, it is quite possible to take into consideration the
quasilocal operators (those with generalized statistics).
They are combined into multiplets with respect to the action
of two quantum groups. The form factors of the fields
which correspond to the highest vectors of the multiplets can
be found using the results of [7]. Including the quasilocal fields
we get exactly the same number of fields in SG as the number of local
fields at the free fermion point. Technically, the consideration of
such fields corresponds to omitting the requirement $\#(k_i)=m-1$ in
the formula (16), rather we have to consider  $\#(k_i)\le m-1$.

Second, in this paper we considered the operators which in the SG
language are identified with $\text{exp}(im\b\phi)$ and their
Virasoro descendents. It is possible to consider the operators of disorder
type: $\text{exp}(i(2m+1)\b\phi/2)$ and their descendents.
These operators have infinitely many form factors even at the free
fermion point. The formula (16) in application to them will
have an additional multiplier:
$$\prod B_j^{-{1\over 2}}\prod A_i $$
\head
6. Discussion.
\endhead

We do suppose that further development of the results of
the present paper will lead to a much better understanding
of the integrable field theory. Let us outline the most promising
directions for the future study.

1. On purpose we did not go into details of symmetry: it has to
be done separately. However, let us outline briefly the
structure. It is known that SG allows the algebra of nonlocal
charges $U_q(\widehat{sl}(2))$ for $q=-\tau ^2$ [2]. The results of
this paper imply the existence of another hidden algebra
$U_q(\widehat{sl}(2))$ with $q=-1$. This algebra is not a symmetry
of the theory, but it is responsible for counting the local fields.
The canonic transformation with the deformed
matrix of periods adjusts this two algebras or makes transformation from
fields to particles. At the free fermion point the fields and
the particles are the same which is implied by simplicity of
the period matrix at this point. That is why the two algebras also
coincide at the free fermion point.

2. Probably the most adequate point of view to the SG theory is
the following. We have to consider not many fields depending
on the local times $t_{2p+1}$, but one field depending
on the local times $t_{2p+1}$, on the additional times $t_{2p}$
and on the fermions $a,b$ (for $\a$ and $\b$ cycles)
which describe the deformed cycles.
So, the space where SG field lives is much bigger than
usually expected.

3. It is very important that the field counting structure is
independent of the coupling constant. The most important consequence of
the fact is the following one. Even in the classical limit $\xi\to 0$
the structure stays untouched, so, we have to think about classical
interpretation. Certainly, the very meaning of the structure
implies the connection with the dressing symmetries
(see [1] for relevant discussion): it is not
a real symmetry, but it counts different fields. We have no doubts that
there must be a relation to Sato's Grassmanian.

Another interesting
limit corresponds to the asymptotically free quasiclassics: $\xi\to \infty$.
This limit seems to be more interesting than the classical SG one, but
in spite of several attempts the understanding of the situation is not
absolutely clear. Let us explain why this limit is so important.
{}From the results of the present paper it follows that the duality
between particles and fields in SG model is related to the
duality between the deformed cohomologies and homologies of
hyperelliptic surface. The limit  $\xi\to \infty$ is classical
for this picture: this is the limit where we get usual surfaces with
their homologies and cohomologies.

4. Returning to the free fermion point one sees that the
introduction of the variables $\a _i$ is nothing but a fancy
way of describing the spinor structure of fermions (similar to
the screening operators technics). It must be possible to show that the
formula (16) solves a version of Thirring equation of motion.
Important thing is that this way of writing the
equations of motion must be free of divergencies:
all the renormalizations are absorbed into the period
matrix.

5. The technics of this paper can be generalized in order to
describe the form factors for the massless flows [17] from the principal
chiral field into WZNW theory.

6. Finally, it is amusing that the formula (16) looks as
a $\tau$-function of some classical integrable equation.
Can such an equation be found, and what is its meaning
for the SG theory?

\head
Acknowledgements.
\endhead
I would like to thank O. Babelon, D. Bernard,
S. Lukyanov, T. Miwa and A. Zamolodchikov for discussions.

\end